\documentclass{agujournal}

\journalname{Journal of Geophysical Research - Space Physics}

\usepackage{hyperref}
\hypersetup{
    colorlinks=true,
    linkcolor=blue,
    urlcolor=cyan,
}

\begin{document}

\title{The influence of magnetic field topology and orientation on the distribution of thermal electrons in the Martian magnetotail}%
%
\authors{Murti Nauth\affil{1}, Christopher M. Fowler\affil{1}, Laila Andersson\affil{2}, Gina A. DiBraccio\affil{3}, Shaosui Xu\affil{1}, Tristan Weber\affil{2}, David Mitchell\affil{1}}

\affiliation{1}{Space Sciences Laboratory, University of California, Berkeley, California, USA}
\affiliation{2}{Laboratory for Atmospheric and Space Physics, University of Colorado, Boulder, Colorado, USA}
\affiliation{3}{Laboratory for Planetary Magnetospheres, NASA Goddard Space Flight Center, Greenbelt, Maryland, USA}

\correspondingauthor{Murti Nauth}{murtinauth@berkeley.edu}

\begin{keypoints}
\item The local magnetic topology, flaring angle, and location of the crustal fields influence the thermal electron structure in the magnetotail
\item Close to the planet, thermal electrons are most likely observed just outside the optical shadow in the magnetotail, peaking at $\sim$35\%
\item Strong crustal fields on the dayside show thermal electrons are 10-20\% more likely to be observed in the central tail, than the nightside

\end{keypoints}

\begin{abstract}

Thermal (<1 eV) electron density measurements, derived from the Mars Atmosphere and Volatile Evolution's (MAVEN) Langmuir Probe and Waves (LPW) instrument, are analyzed to produce the first statistical study of the thermal electron population in the Martian magnetotail.
Coincident measurements of the local magnetic field are used to demonstrate that close to Mars, the thermal electron population is most likely to be observed at a cylindrical distance of $\sim$1.1 Mars radii ($R_{M}$) from the central tail region during times when the magnetic field flares inward toward the central tail, compared to $\sim$1.3 $R_{M}$ during times when the magnetic field flares outward away from the central tail.
Similar patterns are observed further down the magnetotail with greater variability.
Thermal electron densities are highly variable throughout the magnetotail; average densities are typically $\sim$20-50 cm$^{-3}$ within the optical shadow of Mars and can peak at $\sim$100 cm$^{-3}$ just outside of the optical shadow.
Standard deviations of 100\% are observed for average densities measured throughout the tail.
Analysis of the local magnetic field topology suggests that thermal electrons observed within the optical shadow of Mars are likely sourced from the nightside ionosphere, whereas electrons observed just outside of the optical shadow are likely sourced from the dayside ionosphere.
Finally, thermal electrons within the optical shadow of Mars are up to 20\% more likely to be observed when the strongest crustal magnetic fields point sunward than when they point tailward.

\end{abstract}
\section{Introduction}
Mars lacks an intrinsic dipole magnetic field and the interaction between the planet's atmosphere and solar wind results in the formation of a partially induced magnetosphere that acts to stand off and deflect the supersonic solar wind flow around the planetary obstacle \citep{luhmann1991, brain2003, Bertucci2011}.
The interplanetary magnetic field (IMF) drapes about the planet's dayside, leading to a formation of a flared ``wake'' behind the planet, similar to Venus, comets, and other unmagnetized bodies (e.g. \citealt{vaisberg1986martian,luhmann1991,zakharov1992,Israelevich1994}).
Analysis of the structure and dynamics of the magnetotail region inform us of the interaction between the planet and the supersonic solar wind.
Planetary ions have also been observed traveling tailward within the wake \citep{lundin1992}, and thus understanding magnetotail dynamics is essential to understanding the global structure of the induced Martian magnetosphere and ion loss to space.

Early Mars orbiters, such as the Phobos spacecraft, demonstrated that the Martian tail is a two-lobe structure whose polarity and orientation depends on the upstream solar wind orientation \citep{Yeroshenko1990}, similar to the Venusian magnetotail \citep{vaisberg1986martian}.
Data from the Mariner 4, Mars 2, 3, 5, Phobos and Mars Global Surveyor (MGS) spacecraft missions show the shape and extent of the Martian magnetotail is highly variable (e.g. \citealt{Slavin1991,Vignes2000}), and the flaring angle of the tail depends on the upstream solar wind pressure \citep{Zhang1994}.

Later spacecraft, including MGS, Mars Express (MEX) and the Mars Atmosphere and Volatile EvolutioN (MAVEN) mission, have enhanced these early observations, demonstrating that a current sheet separates both tail lobes, as expected for an induced magnetotail produced by the draping of the IMF about the dayside of the planet \citep{Ferguson2005}.

\citealt{Halekas2006} have also shown that MGS global current sheet crossings at 400 km altitude occur everywhere (including the dayside) except over the strong Martian crustal magnetic fields.
Such crossings occur at locations that vary with Mars season and upstream IMF orientation.
The magnetotail region has been observed to be at times highly dynamic.
Evidence of magnetic reconnection occurring in the tail region has been reported using both MGS and MAVEN data (e.g. \citealt{dibraccio2015,Halekas2009,Eastwood2012,Harada2015}), while the repetitive loading and unloading of magnetic flux in the magnetotail region is akin to sub-storm activity within intrinsic magnetospheres (e.g. \citealt{dibraccio2015}).
\citealt{DiBraccio_twist} also demonstrated that the magnetotail lobes typically exhibit a 45 degree twist relative to upstream IMF orientations, a characteristic thought to be strongly influenced by dayside reconnection between the draped IMF and planetary crustal magnetic fields.
\citealt{xu_2020} compared the tail topology determined from magnetohydrodynamic (MHD) simulations to that inferred from MAVEN data and how each topology responds to the upstream IMF orientation.
Their results support the hypothesis that magnetic reconnection between crustal magnetic sources and the solar wind is responsible for the observed twist in Mars's tail lobes.

Planetary ions have been observed in the magnetotail by multiple spacecraft, and have been used to infer the existence of various plasma acceleration mechanisms and subsequent plasma outflow from Mars.
Ion data from the Analyzer of Space Plasmas and Energetic Atoms (ASPERA) instruments on Phobos 2 and MEX show highly variable O$^+$ fluxes, particularly in the central tail region, and these fluxes are likely driven by a variety of ion acceleration mechanisms.
Some processes are similar to those that exist in the terrestrial magnetotail and others are different \citep{lundin1989,kallio1995,lundin2004,federov2006,Lundin2008}.
Magnetic field and suprathermal electron measurements from the MGS spacecraft show that detached magnetic structures intermittently exist in the Martian tail region and contain planetary plasma.
Such structures are thought to be caused by the stretching of crustal magnetic fields via their interaction with the solar wind in the tail region, until magnetic reconnection occurs, detaching part of the crustal field, which is then convected down the tail region \citep{brain2010}.

Initial MAVEN estimates of planetary ion escape rates for ions with energies > 25 eV from Mars generally agree with earlier studies, demonstrating that tailward ion escape in the magnetotail region can contribute significantly to the total escaping flux from the planet \citep{brain2015, Dong_2015}.
MAVEN observations of enhanced ionospheric electron temperatures above the exobase region suggest that ambi-polar electric fields can be an important ion acceleration mechanism at Mars, providing up to $\sim$1 eV of energy to planetary ions \citep{xu_field_aligned,col_2019,akbari_2019}.
Such energy is close to the escape energy for heavy ions to overcome Mars' relatively weak gravitational potential and drive ion outflow on open magnetic field lines that connect the ionosphere to the solar wind.
Numerical simulations have explicitly shown that ion outflow can be significantly enhanced by such ambi-polar electric fields at Mars \citep{ergun2016}.
MHD modelling of the Martian magnetosphere, for example, has shown that draped field lines exist in abundance in the magneotail region \citep{ma2002,ma2004,fang2018}.

Instrument limitations make it difficult to study the thermal (< few eV energy) plasma environment in the Martian magnetotail with electrostatic analyzers.
The arrival of the MAVEN spacecraft at Mars in September 2014 \citep{jakosky2015} presented an additional method to observe the planet's low energy thermal plasma environment, in the form of the Langmuir Probe and Waves (LPW) instrument.
LPW measurements enable a derivation of the local, thermal (<1 eV) electron density, and this study presents, to our knowledge, the first statistical analysis of the thermal electron population in the Martian magnetotail.
We demonstrate that the location at which thermal electrons are observed within the magnetotail is highly dependent upon the orientation of the local magnetic field, whether it connects to the day or night side ionosphere, and the location of the strongest crustal fields.
Section \ref{data_big} describes the dataset and analysis methods utilized in this study; results are presented in Section \ref{results} and discussed in \ref{discuss}. We conclude in Section~\ref{conclude}.


\section{Data and Methodology} \label{data_big}

\subsection{Instruments}
The MAVEN spacecraft entered into an elliptical orbit about Mars in September 2014 \citep{jakosky2015}.
The orbit precesses such that MAVEN samples all local times, longitudes, and latitudes.
The rate of orbital precession results in coverage of the Martian magnetotail roughly every 3-4 months, for about one month each time.
The data used in this study are taken from such ``tail seasons.''
Data from several instruments were utilized in this study.
These include the Langmuir Probe and Waves (LPW) \citep{andersson2015lpw}, Magnetometer (MAG) \citep{connerney2015mag}, and Solar Wind Electron Analyzer (SWEA) \citep{mitchell2016swea}.

LPW consists of two Langmuir probes, each mounted on the end of $\sim$7m booms and separated by an angular distance of 110 degrees.
The instrument can operate as a Langmuir probe (``when in LP mode'') or as an electric field instrument (``when in waves mode'').
During LP mode, the Langmuir Probes measure (alternating in time) the current and voltage (I-V) characteristics of the local plasma environment.
The density and temperature of local thermal (< 1eV) electrons and the spacecraft potential are derived from the I-V curve by using an enhanced fitting method \citep{ergun2015}.
During waves mode, the instrument measures---as a time series--- the potential difference between the two sensors, from which electric field power spectra are derived \citep{andersson2015lpw, fowler2017}.
Only data obtained from LPW's LP mode were used in this study.

The LPW instrument is designed to measure thermal (<0.1 eV), high density (>1000's cm$^{-3}$) dayside ionospheric plasma.
However, there are times when the local thermal electron density is low ($\lesssim 15 $cm$^{-3}$) and the electron temperature is relatively high (approaching 1 eV or larger).
During such conditions the LPW instrument is operating beyond its design specifications and at times may not be capable of measuring a significant thermal electron current.
Additionally, if the probe and/or spacecraft are sunlit, photoelectron currents can dominate the thermal electron current, when the thermal electron density is below 10-15 cm$^{-3}$.
Consequently, when the thermal electron density is below $\sim$15 cm$^{-3}$, derived thermal electron densities can be close to or equal to zero, and have large uncertainties associated with them.
These limitations should be kept in mind when interpreting the results of this study.


The local three-dimensional magnetic field vector is measured by the MAG instrument at a sampling rate of 32 Hz.
The instrument consists of two fluxgate magnetometers which allow for hardware redundancy, calibration, and removal of the spacecraft generated magnetic fields.
The magnetic field is measured to an accuracy of about 0.01 nT \citep{connerney2015mag}.

The SWEA instrument is an electrostatic top-hat analyzer that measures electron fluxes within the 3 eV to 5 keV energy range.
The instrument has a field of view of 360$^\circ$ $\times$ 120$^\circ$ provided by electrostatic deflectors and an energy resolution ($\Delta E/E$) of 17\% \citep{mitchell2016swea}.
SWEA operates at a cadence of 2-4 seconds.
As described in section \ref{methods}, SWEA shape parameters were used to determine magnetic field topology.

In this study, the data were analyzed in the Mars Solar Orbital (MSO) coordinate system, where the x-axis points from Mars to the Sun, the y-axis is anti-parallel to Mars' instantaneous orbital velocity, and the z-axis completes the right-handed coordinate system.

\subsection{Methodology} \label{methods}
The goal of this study is to investigate the effects of the local magnetic field topology on the spatial distribution of the thermal electron population in the Martian magnetotail.
To this end, we analyze MAVEN data from periods when the spacecraft was present in the magnetotail region between 01-01-2015 to 12-31-2019.

Every LPW observation for which the I-V curve fitting technique detected a thermal electron density in the magnetotail was binned according to its observed location within the magnetotail region and the associated local magnetic field orientation and topology.
This is further described below.
The magnetotail region was defined as all post-terminator locations (X (MSO) < 0 ) with altitudes greater than 600 km.
All LPW I-V sweep measurements were then binned into two spatial regions of the tail---called regions X1 and X2---and separated by their associated local magnetic field orientation and topology.
Region X1 enclosed $-1.5 R_{M} <X< 0 R_{M}$ and region X2 enclosed $-3 R_{M} <X< -1.5 R_{M}$, where $R_{M} = 1$ Mars radius (3396 km).
Splitting the tail region into two regions provides the opportunity to observe the evolution of the spatial distribution of thermal plasma down the tail.
However, spatial coverage at this time in the MAVEN mission means the tail region could not be split into additional regions while maintaining adequate sample numbers.

\begin{figure}[!ht]
    \centering
    \includegraphics[ clip=true, width= \textwidth] {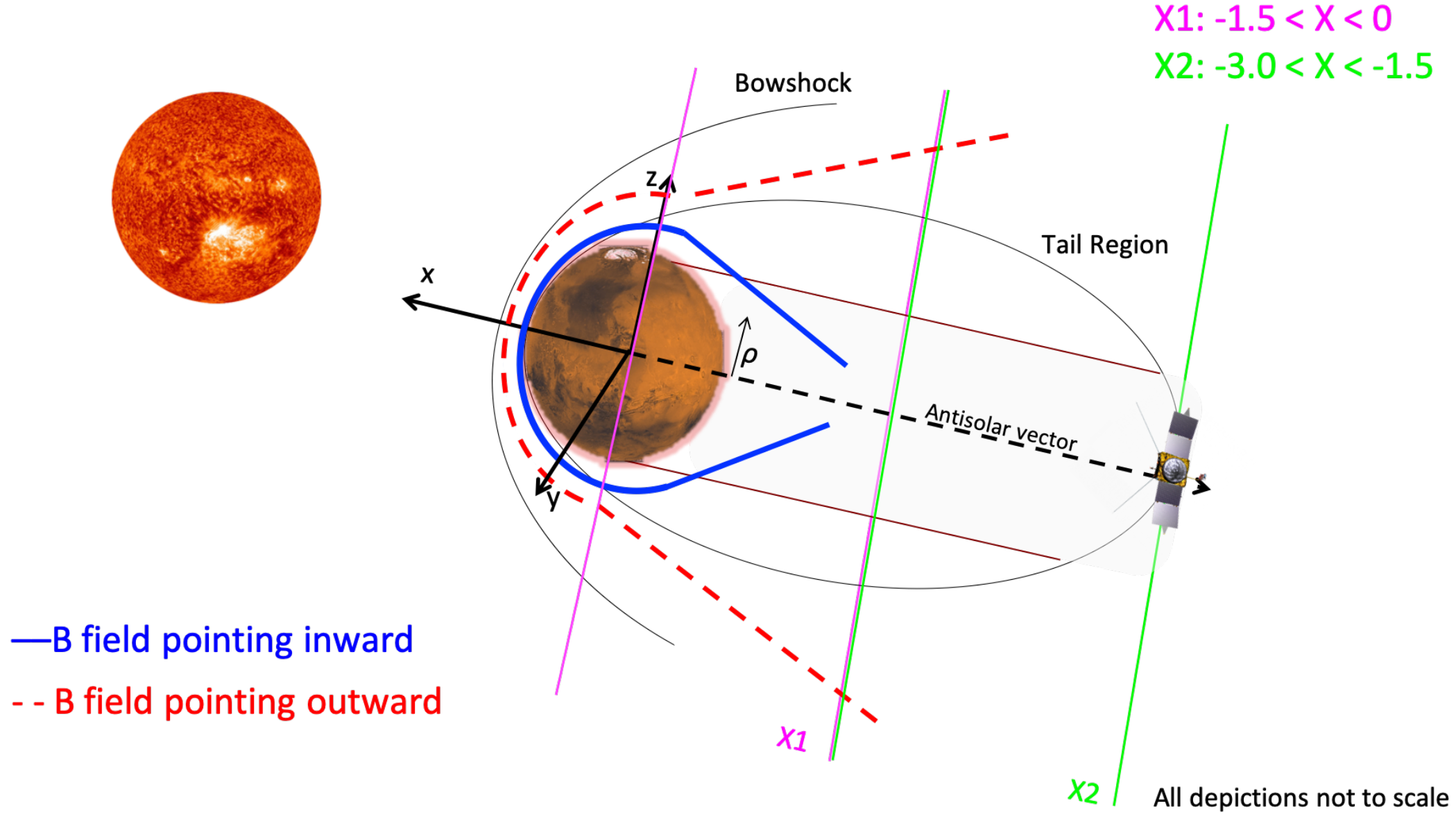}
    \caption{Cartoon depicting the tail region. The cylindrical distance from the center of the tail is given by $\rho=\sqrt{ Y^2 + Z^2}$ in the MSO coordinate system and reported in units of Mars radii ($R_M$).
    The relative angle between the magnetic field vector and the anti-solar vector determines the flare orientation, as described in section \ref{methods}.}
    \label{fig_cartoon}
\end{figure}

Data were also analyzed with respect to the parameter $\rho$, the cylindrical distance of MAVEN from the center of the tail region, defined as $\rho = \sqrt{Y^2 + Z^2}$, in the MSO coordinate system and reported in units of Mars radii ($R_M$).
To first order, $\rho < 1\ R_M$ lies within the planet's optical shadow and $\rho > 1\ R_M$ lies outside the optical shadow (ignoring atmospheric effects).
These constraints ensured MAVEN was sampling the magnetotail region and not the nightside ionosphere, which is typically observed below altitudes of 600 km (e.g. \citealt{fowler2015}).
A cartoon diagram showing regions X1 and X2, and $\rho$ is shown in Figure \ref{fig_cartoon}.

Two distinct datasets were utilized in this study.
The first are dubbed ``measurement points'' which represent all measurements made by LPW regardless of whether sufficiently large currents were observed such that a density could be derived from the measured I-V curve or not.
The second, ``derived data points,'' represent measurements from which densities were derivable, corresponding to times when the thermal electron density was larger than $\sim$15 cm$^{-3}$.
Dividing the number of derived data points by the number of measurement points provides a percentage of how often a density can be derived from the measured I-V curves.
The 4-year-long dataset consisted of $10^5$ measurement points in the tail region.
Of these $10^5$ measurement points, $\sim6\times10^4$ are derived data points.
Since LPW is sensitive down to densities of $\sim$15 cm$^{-3}$, we may lack measurements of the lowest densities.

Furthermore, each MAG and SWEA measurement were paired to a corresponding LPW measurement in time.
Both MAG and SWEA operate at significantly faster cadences than LPW at high altitudes within the magnetotail region (128~s, compared to 2~s for SWEA and 32~s$^{-1}$ for MAG).

The flaring angle of the magnetic field was calculated based upon the relative angle between the local magnetic field vector and the anti-solar vector (-X) in the three-dimensional MSO coordinate system.
The dashed red line and solid blue line in Figure \ref{fig_cartoon} depict outward and inward flaring field, respectively (as projected onto the two-dimensional plane).
When combined with LPW density measurements, the flaring angle and topology provide insight into the origin of the electrons measured on the field line, i.e. the nightside vs the dayside ionosphere.
This study utilizes three primary magnetic field orientations: closed, open and draped.

Suprathermal electron energies and pitch-angle distributions measured by SWEA allow for the inference of magnetic topology based on three principles.
First, the presence of a loss cone in one or both field-aligned directions indicates a magnetic field line intersects the collisional atmosphere once or twice.
Thus, the field line is considered open or closed (e.g. \citealt{loss_cone1,loss_cone3, loss_cone2}).
Second, ionospheric photoelectrons are observed in one or both field-aligned directions.
This implies one or both footpoints of a field line are embedded in the dayside ionosphere, corresponding to either open or closed field topology (e.g. \citealt{xu_2017}).
Finally, strong depletion of suprathermal electron flux, called ``suprathermal electron voids,'' signifies closed field lines with both footpoints intersecting the nightside ionosphere (e.g. \citealt{mitchell_reflect, steckiewicz}).
Note, magnetic topology or footpoint(s) of a field line is defined with respect to the suprathermal electron exobase ($\sim160$ km, \citealt{xu_2016}).
If none of the above is observed, then the field line is draped.
The methodology of identifying photoelectrons and loss cones through MAVEN data are described in further detail in \cite{xu_2017} and \cite{loss_cone2}, respectively.
This study infers thermal electron origin (dayside versus nightside ionosphere) based on the magnetic topology identification method described by \cite{ xu_2016, xu_2019}.

In addition to the magnetic field topology in the tail region, the effect of the Martian crustal magnetic fields (e.g. \citealt{acuna1999}) on the distribution of the thermal electrons in the tail was also investigated.
The location of the region which contains the strongest fields (e.g. planetary longitudes of 180 degrees--- \citealt{acuna1999}) with respect to the sub-solar point was analyzed when the thermal electron population was observed.
These ``relative sub-solar longitudes'' are the difference between the position of the strongest crustal fields and the sub-solar point.
Relative sub-solar longitude values close to 0 degrees indicate the strongest crustal fields are pointing sunward and values close to 180 degrees indicate they are pointing tailward.
Similarly, intermediate relative sub-solar longitude values of 90 degrees and 270 degrees, signify the strongest crustal fields are located at dawn and dusk, respectively.

Derived data points and measurement points were further binned based on relative sub-solar longitude bins (0$^{\circ}$ $\pm$ 45$^{\circ}$, 90$^{\circ}$ $\pm$ 45$^{\circ}$, 180$^{\circ}$ $\pm$ 45$^{\circ}$ and 270$^{\circ}$ $\pm$ 45$^{\circ}$), corresponding to strong crustal fields located at noon, dawn, midnight, and dusk.
This was done regardless of whether the measurements belong to region X1 or X2 to ensure adequate sampling statistics.

\subsection{Example Data}

\begin{figure}[!ht]
    \centering
    \includegraphics[width=\textwidth]{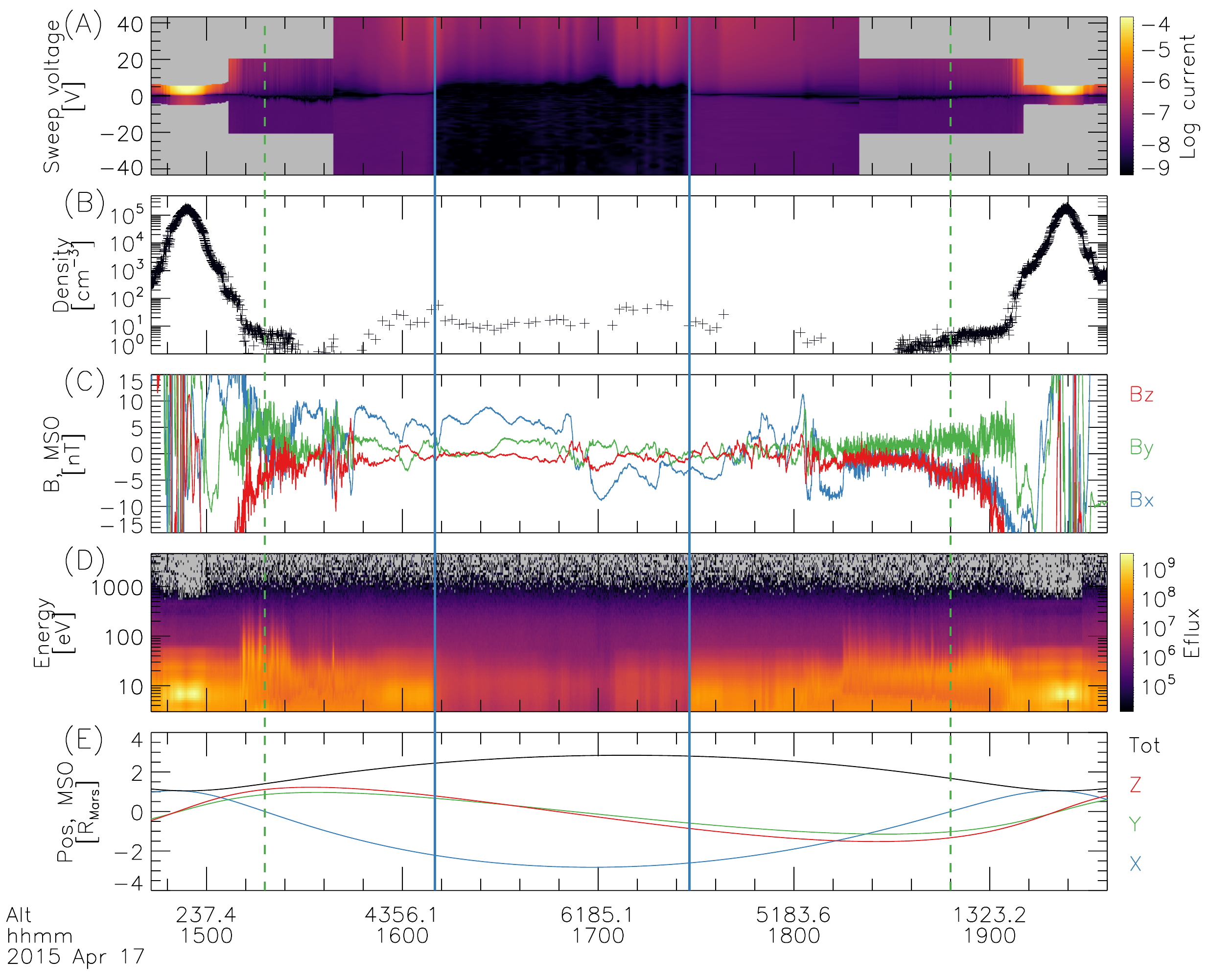}
    \caption{One MAVEN orbit while the spacecraft was present in the tail region. The top panel (A) shows LPW I-V sweeps with voltage on the Y-axis and the log of the absolute value of the current as color. Log current has units of $\log_{10}(A)$. Thermal electron densities derived from the I-V sweeps are in panel B. The 3D MAG data in the MSO frame are in panel C. The SWEA suprathermal electron spectrum is shown in panel D, where ``Eflux'' has units of $\frac{eV}{eV~s~sr~cm^2}$.
    Panel E shows MAVEN position and altitude (black line) in the MSO frame; altitude values (km) are printed underneath. The dashed green lines mark altitudes of 600 km and the solid blue lines enclose the optical shadow, which extends to $\rho \sim 1$.}
    \label{fig_orbit}
\end{figure}

Example MAVEN data are shown as a time series in Figure \ref{fig_orbit}.
The time series data span a period of 6 hours, which includes two periapsis passes at $\sim$15:00 and 19:30 UTC.
These periapsis passes are located on the dayside of the planet and are characterized by typical peak electron densities $\le$ ${ 10^5}$~cm$^{-3}$ (Figure \ref{fig_orbit}B).
Apoapsis occurs just after 17:00~UTC in the magnetotail region.
MAVEN crosses into the planet's optical shadow between 16:00 and 17:30, as enclosed by the two vertical solid blue lines.
The effect of the optical shadow is observed as a reduction of photoelectron-current in the LPW I-V sweeps (negative voltages, panel A) and a reduction of SWEA suprathermal electron flux (due to negative spacecraft potentials within the optical shadow, energies less than $\sim$20 eV, panel~D).

MAVEN crosses the magnetotail current sheet at about 17:00, as indicated by the change in sign of the magnetic field's $B_{X}$ component (panel C;  e.g. \citealt{dibraccio2015}).
Thermal ionospheric electrons are observed in the magnetotail region, as shown in panel B.
The vertical dashed green lines mark an altitude of 600 km, the minimum altitude above which data were analyzed in this study.
At high altitudes, the LPW data are measured at lower time cadences.
This results in fewer I-V sweeps and density measurements at higher altitudes, as shown in panels A and B.

\subsection{Removal of LPW Photoelectron signatures from LPW Densities} \label{photoelec}
\begin{figure}[!ht]
    \centering
    \includegraphics[width=\textwidth]{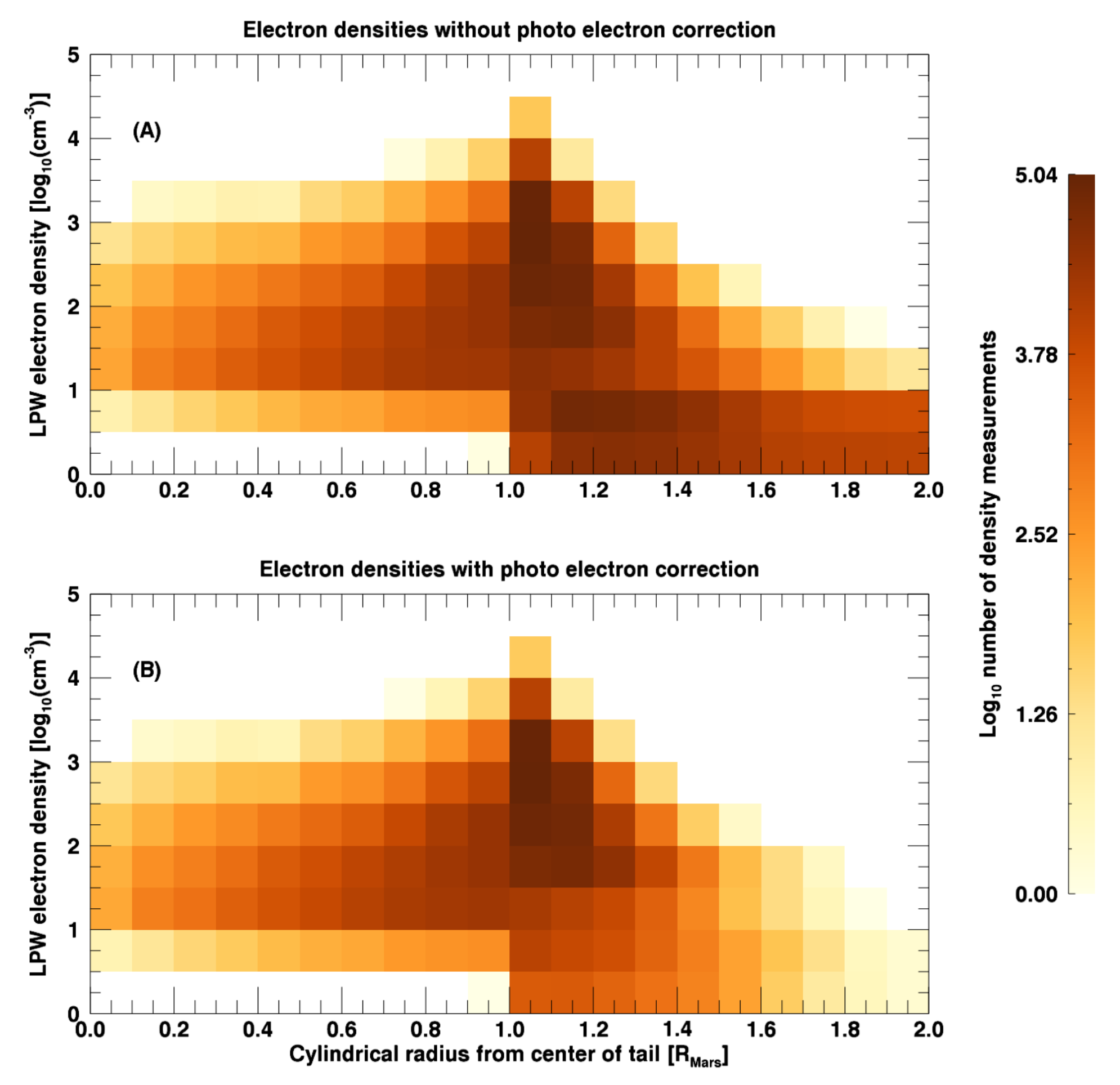}
    \caption{The number of LPW derived data points as a function of radial distance ($\rho$) from the center of the tail and their respective derived densities. Panel A shows the measurement points without the photoelectron correction and panel B shows the measurement points with the 10~cm$^{-3}$ photoelectron current correction applied.}
    \label{fig_photos}
\end{figure}

During LP operation mode, the Langmuir Probe sensor measures electrical current from a variety of sources, including thermal electrons, ions and photoelectrons emitted by the sensor when in sunlight.
In sunlit conditions, the photoelectron current emitted by the LPW sensors can dominate the collected current, when the local thermal electron density is small.
This results in a ``background'' derived thermal electron density of $\sim$10 cm$^{-3}$.
An example of this background density influenced by photoelectrons is shown in Figure \ref{fig_photos} (A).
The panel shows the number of LPW measurement points as functions of radial distance ($\rho$) from the center of the tail and their respective derived densities.
Radial distances less than 1 $R_{M}$ denote measurements made within Mars' optical shadow, while measurements made at radial distances greater than 1 are made in sunlight.
Within the optical shadow region ($\rho <1\ R_M$), an ``artificial cutoff'' at densities below 10$^{0.5}$ cm$^{-3}$ demonstrates the lower measurement limit of the LPW instrument, when photoelectrons are not present.
Outside of the optical shadow ($\rho > 1\ R_M$), there is a clear ``background'' of measured densities at $\sim$10 cm$^{-3}$, which are a result of photoelectron currents dominating the collected current.
The LPW instrument team previously investigated such cases in detail (not shown here) and confirmed that the photoelectron ``background current'' is equivalent to a density of about 10 cm$^{-3}$.

We correct the LPW thermal electron densities for this background photoelectron current before performing the analysis mentioned in Section \ref{methods}.
Derived densities measured in sunlight ($\rho > 1\ R_{M}$) have 10 cm$^{-3}$ subtracted from their derived values, and the results of this correction are shown in Figure \ref{fig_photos} (B).
The format is the same as for panel~A. For radial distances greater than 1, the transition to density values below 10 cm$^{-3}$ is now much smoother, indicating that this background correction is successful.
We note here that the photo electron current correction of 10 cm$^{-3}$ is only significant when derived thermal electron densities are less than $\sim$50 cm$^{-3}$.
As shown in Figure \ref{fig_orbit}, the majority of derived density values are greater than 100 cm$^{-3}$, and this correction is negligible for most data points.

\section{Results} \label{results}

\subsection{Statistics and the Influence of the Magnetic Field Flaring Angle}

\subsubsection{Region X1: -1.5 $R_M$ < X < 0 $R_M$}
\begin{figure}[!ht]
  \centering
    \includegraphics[width=\textwidth]{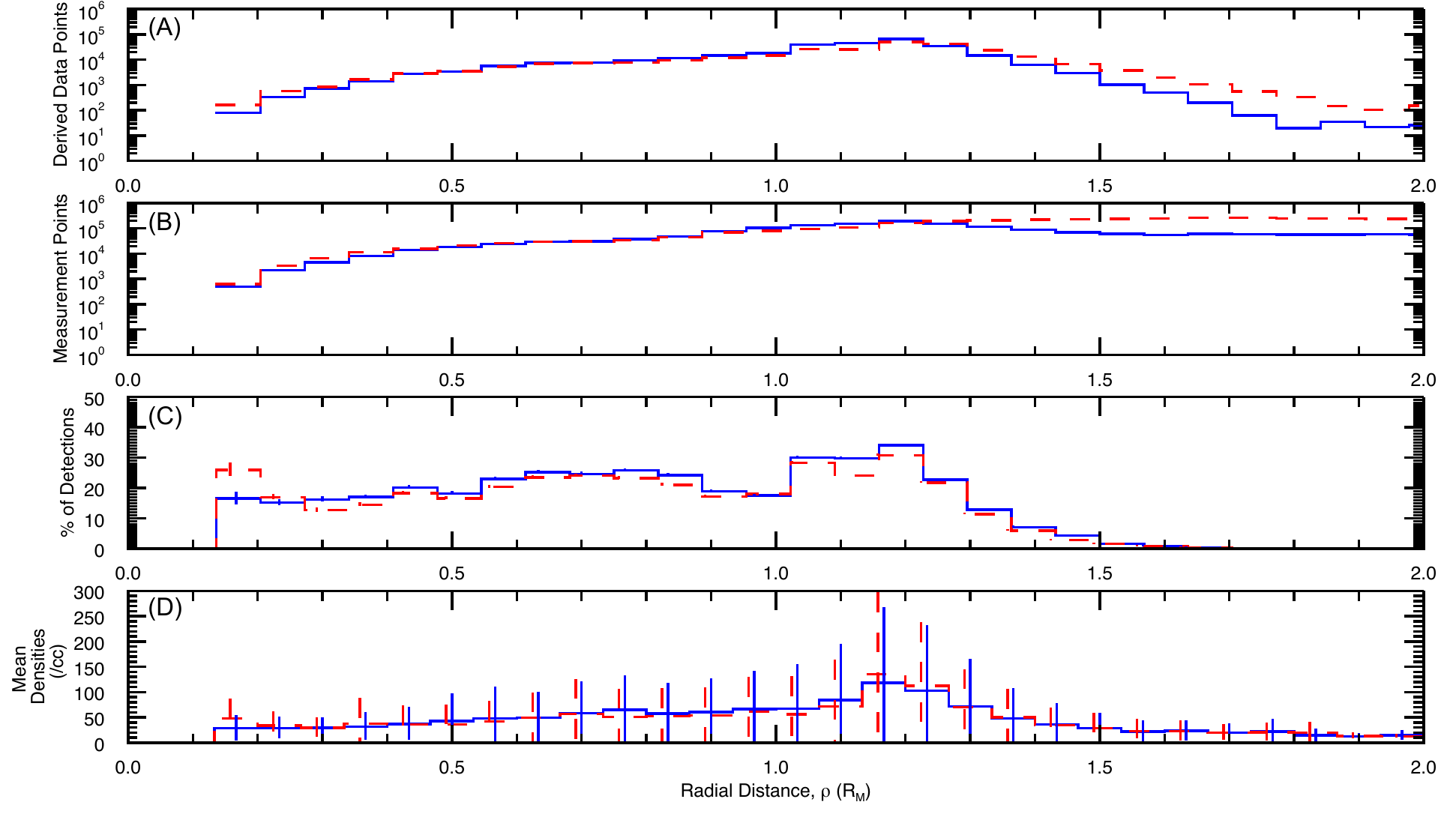}
    \caption{The number of derived data points (A), the number of measurement points (B); the likelihood of observing a density, (panel A / panel B), (C); and the mean densities (D) for all $\rho$ in \textbf{region X1}. The blue line indicates the magnetic field flares inward and the dashed red line indicates the field flares outward. Error bars in C are counting errors and error bars in D are the standard deviations of each bin.}
    \label{fig_x1_all}
\end{figure}

The statistics of the data and measurement points in region X1 of this study are presented in Figure \ref{fig_x1_all}.
Plotted against $\rho \ (R_M)$, panel A shows the number of derived data points, panel B shows the number of measurement points, panel C shows the percent likelihood of observing a density, i.e. panel A divided by panel B, and panel D shows the mean densities for times when densities are observed, i.e. panel A.
The solid blue line indicates an inward flaring field and the dashed red line indicates an outward flaring field.


The LPW observations analyzed in this study show that thermal electrons are observed with relatively high frequency in the magnetotail region of Mars, with typical likelihoods of $\sim$30\% in region X1 close to Mars, $0.5 \ R_M < \rho < 1\ R_M$,  (Figure \ref{fig_x1_all}C).
Thermal electrons are most likely to be observed just outside the optical shadow ($\rho$ slightly greater than 1~$R_M$), and this likelihood then decreases with increasing $\rho$.
The near imperceptible error bars in \ref{fig_x1_all}C represent the counting errors (i.e. $\sqrt N$, where $N$ is the number of derived data points).

Figure \ref{fig_x1_all}D shows mean electron density (when electrons are observed, i.e. Figure~\ref{fig_x1_all}~A) peaks at $\sim$100 cm$^{-3}$ just outside of the optical shadow, $\rho\sim1.2\ R_M$.
Densities within the optical shadow ($\rho < 1\ R_M$) are typically a few 10's to 50 cm$^{-3}$.
The error bars in panel D show the standard deviation of each bin's measurements.
They demonstrate high measurement variability, with standard deviations of $\sim100\%$ in most bins.

The red and blue lines in Figure \ref{fig_x1_all} show that the magnetic field orientation does not significantly affect the likelihood of LPW observing electrons in region X1 (panel~C), nor does it affect the observed average densities in each bin (panel~D).

\subsubsection{Region X2: -3 $R_M$ < X < -1.5 $R_M$}

\begin{figure}[!ht]
    \centering
    \includegraphics[width=\textwidth]{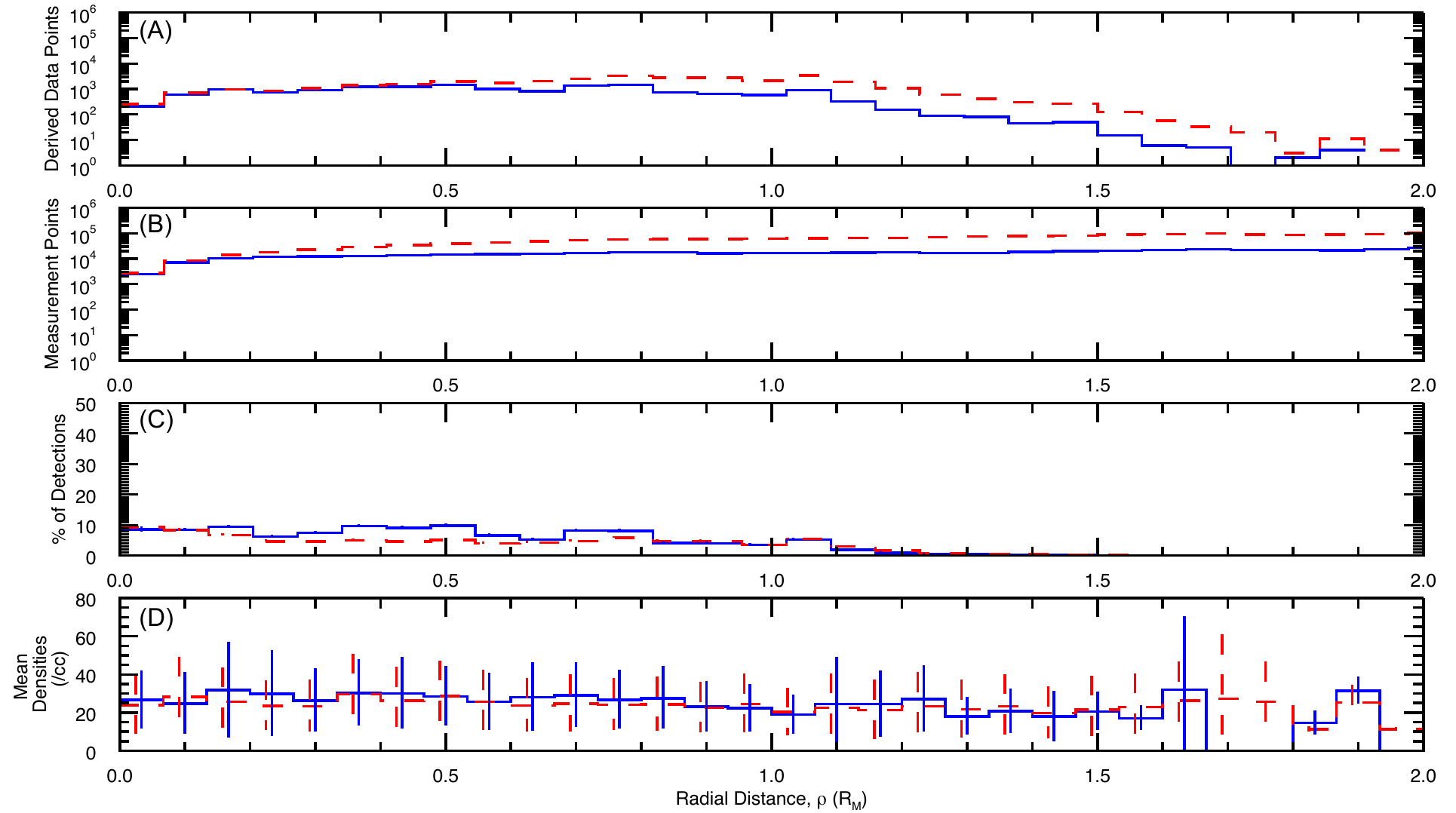}
    \caption{The number of derived data points (A), the number of measurement points (B); the likelihood of observing a density, (panel A / panel B), (C); and the mean densities (D) for all $\rho$ in \textbf{region X2}. The blue line indicates the magnetic field flares inward and the dashed red line indicates the field flares outward. Error bars in C are counting errors and error bars in D are the standard deviations of each bin.}
    \label{fig_x2_all}
\end{figure}

The statistics for region X2 are shown in Figure \ref{fig_x2_all}, which shows distinct differences to region X1.
The likelihood of observing thermal electrons is roughly constant for $\rho < 1$, compared to a decreased likelihood for $\rho < 0.5$ in region X1.
Average densities are typically smaller in X2 than X1 (not exceeding $\sim$30 cm$^{-3}$ on average), and are roughly constant as $\rho$ increases.

In general, a greater number of measurement and derived data points occur when the field flares outward, suggesting that (as expected) the field tends to be in an outward flaring configuration more often than an inward one, particularly at greater distance down the tail (i.e. region X2).
Within $\rho < 1\ R_M$, densities are more likely to be observed when the magnetic field flares inward (blue line, Figure \ref{fig_x2_all}C), though this difference is small, $\sim$10\%, between outward and inward flaring fields.

\subsection{The Influence of Magnetic Field Topology}
Figure \ref{big_fig} shows the likelihood of MAVEN observing thermal electrons as functions of the magnetic field flaring direction and magnetic topology.
We considered measurement points on each specific topology as a fraction of measurement points on all topologies.
We divide the instances when a thermal electron is observed on a specific topology by the sum of all measurement points for all topologies to make these calculations.
Note that the sum of all panels for a given $\rho$ is less than one hundred percent because the SWEA analysis routine cannot always reliably determine the topology.
These ``unknown'' topologies were excluded in this section of analysis.
Influences such as magnetic field topology and upstream solar wind conditions may drive and perhaps mirror the observed trends.
This study was limited to classifying the spatial distribution of thermal electrons in the magnetotail.
To disentangle additional influences such as the upstream solar wind during times when thermal electrons are present and are not present, would require comprehensive magnetic field topology analysis that is beyond the scope of this study.

\subsubsection{Region X1: -1.5 $R_M$ < X < 0 $R_M$}

Magnetic field topology information in Figure \ref{big_fig} A-D shows that electrons observed just past the optical shadow are approximately equally likely to be on open or closed magnetic field lines in region X1.
Panels A-C show that thermal electrons on an outward flaring field are typically most likely to be observed at the largest values of $\rho$, regardless of magnetic field topology.
An interesting caveat to this includes panels \ref{big_fig} B and G, where for $\rho<1\ R_M$, electrons are most likely to be observed in the central tail even for an outward flaring field.
The cause of this is not immediately clear, but the observations in panels \ref{big_fig}~B and \ref{big_fig}~G suggest that an outward flaring magnetic field has a greater influence on thermal electrons in region X2 compared to X1, whereas the blue and red lines show similar behavior for $\rho<$1 $R_M$.

For an inward pointing field originating from the dayside, open and closed field lines (panels A and C), electrons are most likely observed at and just past the optical shadow ($\rho \sim$1-1.2 $R_M$), observed at a peak of $\sim$25\% of the time.
Electrons on open, inward pointing field lines are observed more frequently for $\rho<1\ R_M$ than electrons on open, outward pointing field lines (panel \ref{big_fig}A).
Electrons are most consistently observed on outward draped field lines at the largest values of $\rho$ (panel \ref{big_fig} E), although they are also observed for a significant fraction of the time on an inward pointing, draped field for $\rho<1\ R_M$.


The magnetic field flaring angle and source region for observed thermal electrons (e.g. day or nightside ionosphere) also strongly influence where such electrons are observed, as shown in Figure \ref{big_fig} A - D.
Electrons observed on open field lines originating from the dayside ionosphere (panel A) are most likely to be observed at $\rho \gtrsim1 \ R_M$, particularly when the field is flared outward.
In contrast, electrons sourced from the nightside ionosphere (panel B) are most likely to be observed at $\rho<1\ R_M$.
Dayside originating electrons observed on closed field lines (panel C) demonstrate similar patterns to those on open field lines (panel A).
The dayside electrons on inward pointing field are most likely to be observed just past the optical shadow at $\rho=1-1.2\ R_M$, while those on an outward flaring field are observed at equal likelihood for all large values of $\rho$.
Electrons observed on closed field lines originating from the nightside ionosphere (panel D) are more likely to be attached to an inward pointing field than outward pointing field.

\subsubsection{Region X2: -3 $R_M$ < X < -1.5 $R_M$}

\begin{figure}[!ht]
    \centering
    \includegraphics[width=\textwidth,height=8.5in]{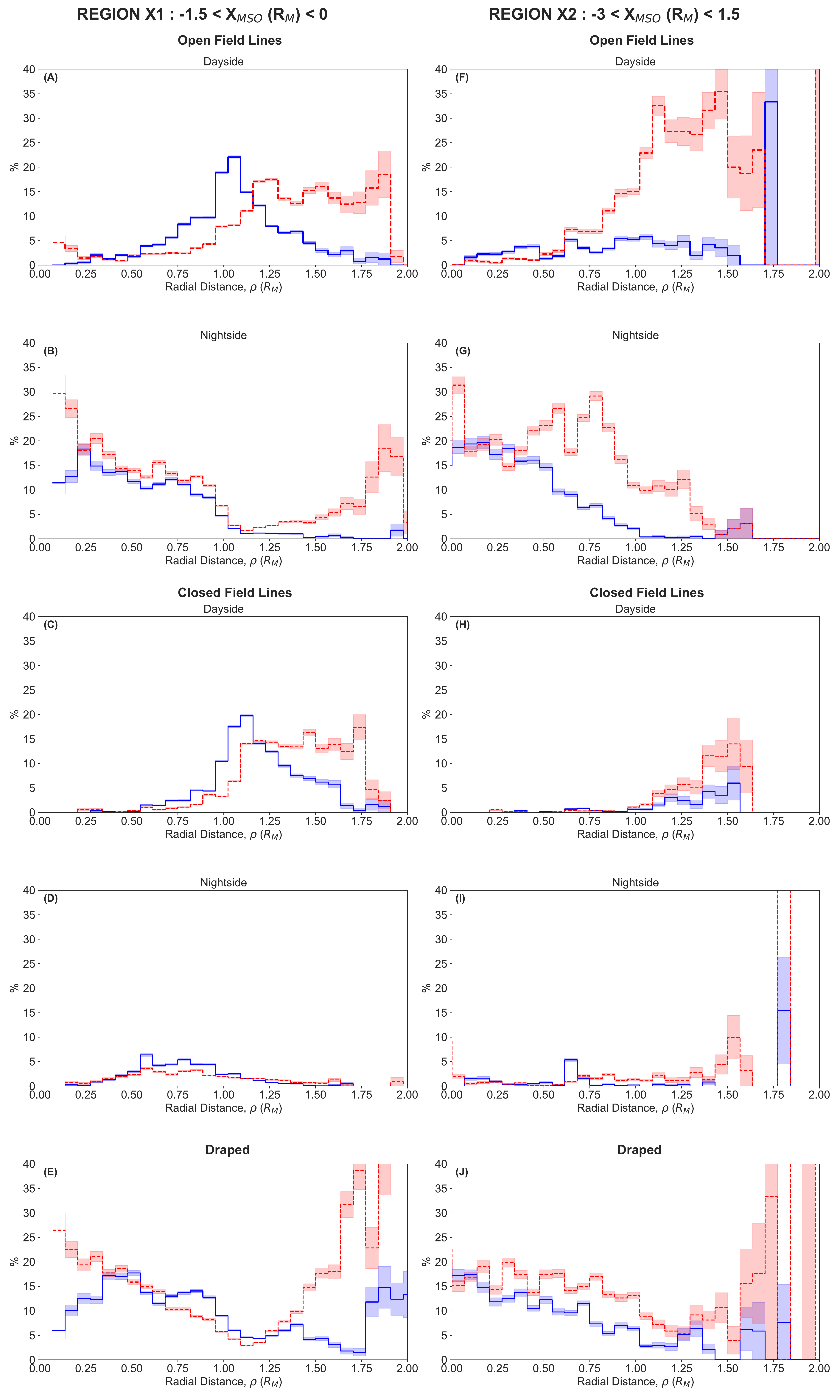}
    \caption{The likelihood of observing electrons in both regions for multiple magnetic field configurations. In \textbf{region X1} the likelihood of observing electrons in open (A,B),
    closed (C,D), and draped (E) field lines is shown. In \textbf{region X2} the likelihood of observing electrons in open (F,G),
    closed (H,I), and draped (J) field lines is shown. The red lines indicate the magnetic field lines flare outward and blue lines indicate the field flares inward. Error bars (shaded regions) represent the counting errors in each bin.}
    \label{big_fig}
\end{figure}

Thermal electrons are most likely to be observed inside of the optical shadow in region X2 ($\rho <1 R_M$), as shown in Figure \ref{fig_x2_all} C.
Panels F - I of Figure \ref{big_fig} demonstrate that the flaring direction of the magnetic field can significantly influence where thermal electrons are observed in region X2.
Panel F shows that electrons are observed with $\sim$30-40\% likelihood at $\rho\gtrsim1.5\ R_M$ when the field flares outward, compared to a peak of $\sim$5\% at the center of the tail when the field flares inward.
Electrons are less likely to be observed on closed field lines in general in region X2 compared to X1 (Figure \ref{big_fig} H and I versus Figure \ref{big_fig} C and D).
This is perhaps not unexpected given that region X2 lies further down-tail away from the closed crustal magnetic fields whose influences are stronger closer to the planet.
The effect of an outward flared field is observed for $\rho\gtrsim1\ R_M$, but the inward and outward flaring directions do not seem to influence the spatial distribution of thermal electrons for $\rho<1\ R_M$.
The presence of an outward flared field in region X2 still leads to a greater likelihood of observation at higher $\rho$.

Electrons observed on draped field lines in region X2 (Figure \ref{big_fig} J) are observed at similar likelihoods (as a function of $\rho$) to region X1 (Figure \ref{big_fig} E).
On draped field lines, thermal electrons are most likely to be observed at small or large values of $\rho$ (panels E and J) for both regions.
It is not immediately clear why the likelihood of observing thermal electrons on draped field increases at the smallest values of $\rho$ in panels \ref{big_fig} E and \ref{big_fig} J.
Comparison with global MHD simulations of the Martian magnetosphere may provide answers to this, although such a study is outside the scope of this work.
In general, thermal electrons are roughly half as likely to be observed in region X2 compared to X1 (Figures~\ref{fig_x1_all} and \ref{fig_x2_all}).

Investigating the source regions for thermal electrons observed in region X2 further demonstrates the clear impact that the magnetic field flaring direction has on the likelihood of observing these electrons.
Figure \ref{big_fig} panels F-I show that thermal electrons are typically more likely to be observed at large $\rho$ when the field flares outward, for field lines connected to the day or nightside ionosphere when compared to thermal electrons in X1.
Inward pointing field connected to the nightside ionosphere (Figure  \ref{big_fig} G) clearly influences where thermal electrons are likely to be observed, with electrons most likely to be observed in the central tail region, similar to region X1.
Figure \ref{big_fig} H and I show that electrons are not often observed on closed field lines in region X2 ($\lesssim15$\% of the time).
When such conditions exist, electrons are more likely to be observed at larger $\rho$ when on outward flared field.

\subsection{The Influence of Strong Crustal Magnetic Fields}\label{crustal-sec}

\begin{figure}[!ht]
    \centering
    \includegraphics[width=\textwidth]{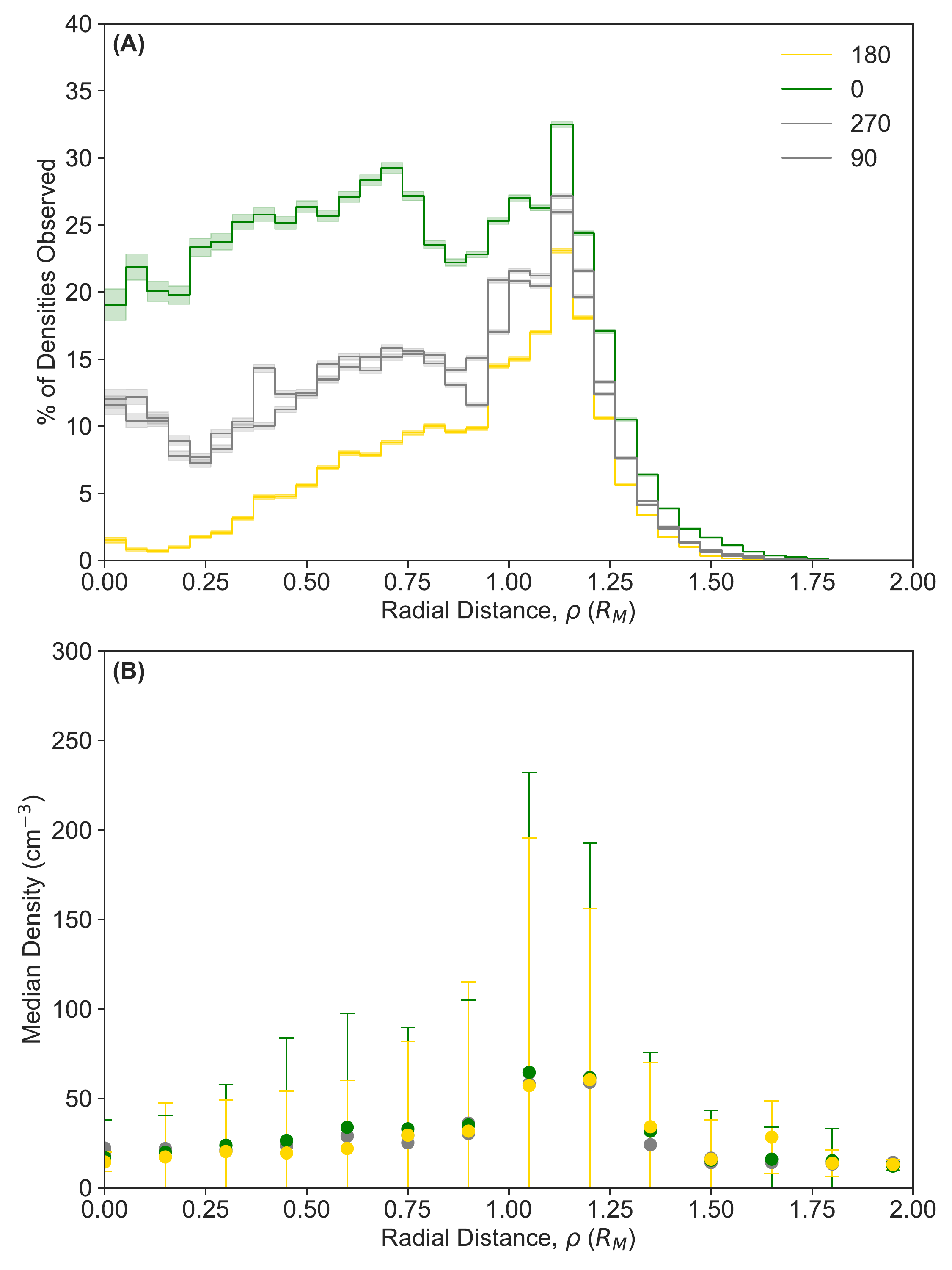}
    \caption{The likelihood of observing thermal electrons based on the location of the strongest crustal fields relative to the sub-solar point (A). The yellow line and dots correspond to the strongest crustal fields pointing tailward (180$^{\circ}$), green line and dots corresponds to the strongest crustal fields pointing sunward ( 0$^{\circ}$), and gray corresponds to dawn and dusk-ward orientation of the crustal fields. Median densities are presented for each crustal field orientation (B).}
    \label{fig_crustal}
\end{figure}

The influence of the location of the strongest crustal magnetic fields on the likelihood of observing thermal electrons in the magnetotail is shown in Figure \ref{fig_crustal}.
For simplicity, we define the strongest crustal fields as those located at 180 degrees planetary longitude (e.g. \citealt{acuna1999}).
The parameter "relative sub-solar longitude" is defined as the longitude difference between the location of the strongest crustal fields (180 degrees planetary longitude) and the sub-solar point.
When the relative sub-solar longitude equals 0 (180) degrees, the strongest crustal fields are on the dayside (nightside) of Mars pointing sunward (tailward).
Values of 90 and 270 degrees denote dawn and dusk locations (90 $\pm$ 45 and 270 $\pm$ 45 degrees, respectively).
Figure \ref{fig_crustal} A shows the likelihood of LPW observing thermal electrons, while \ref{fig_crustal} B shows the median densities observed; both as a function of $\rho$.
The green (yellow) line denotes when the relative sub-solar longitude equals 0 $\pm$ 45 degrees (180 $\pm$ 45 degrees).
The gray lines denote dawn and dusk locations.
Data are combined for regions X1 and X2 to ensure adequate sampling statistics.

The location of the strongest crustal fields drastically alters the location and likelihood of observing thermal electrons, despite showing similar median densities.
The peak likelihood of observation occurs at $\rho\sim1.2\ R_{M}$ for each relative sub-solar longitude bin; however, the distributions vary significantly.
When the strongest crustal fields point sunward (0 degrees, green line in Figure \ref{fig_crustal}A), LPW observes a thermal electron density 10-20\% more often than when the strongest crustal fields point tailward (180 degrees, yellow line in Figure \ref{fig_crustal} A), within the optical shadow for $\rho < 1.2\ R_{M}$.
Beyond $\rho \sim 1.2\ R_{M}$, thermal electrons are $\sim$5\% more likely to be observed when the strongest crustal fields point sunward compared to when they point tailward.

\section{Discussion} \label{discuss}
Thermal electron densities are on average factors of $\sim$2-5 larger in region X1 compared to region X2, depending upon the $\rho$  values being considered.
These electrons originate from the planetary ionosphere, and their presence in the magnetotail is suggestive that they may be contributing to ionospheric escape.
One possible explanation for the differences between the two regions is that, as thermal electrons move tailward, they can be accelerated to energies greater than $\sim$1 eV, and subsequently cannot be measured by the LPW instrument.
Various ion acceleration mechanisms are known to act in the Martian magnetotail (for example, ambi-polar fields \citep{collinson_2015, ergun2015_b, loss_cone3}, $J\times B$ forces \citep{halekas2017}, and magnetic reconnection \citep{Harada, harada2017}).
It is not unlikely that a range of mechanisms also exist that accelerate electrons in the tail region.
Such suprathermal electrons can be measured by the SWEA instrument on board MAVEN, and we leave it to future work to examine the specific electron acceleration mechanisms active within the magnetotail region at Mars.

The local magnetic field direction plays an important role in influencing where thermal electrons are likely to be observed, regardless of distance down the tail.
When the local magnetic field flares outward compared to inward (red vs blue lines, Figure \ref{big_fig}), thermal electrons are typically more likely to be observed at $\rho > 1\ R_M$ for all magnetic field topologies.
This behavior can be explained by the fact that the magnetotail plasma environment is collisionless and electron motion is subsequently dominated by the local magnetic field direction.

The ionospheric source regions to which the local magnetic fields connect also play an important role in driving the spatial distribution of observed thermal electrons.
When thermal electrons are observed on open or closed magnetic field lines, they are sourced primarily from the dayside ionosphere when observed outside of the optical shadow ($\rho >1\ R_M$), while they are sourced primarily from the nightside ionosphere when observed within the optical shadow ($\rho <1\ R_M$).
This behavior is observed in regions X1 and X2 (Figure~\ref{big_fig}~A-D and F-I).
Such a dependence on source region demonstrates that open and closed magnetic fields anchored in the dayside ionosphere “drape” across the terminator into the magnetotail region, such that ionospheric thermal electrons are able travel along these field lines into the magnetotail region.
This behavior likely explains why the largest thermal electron densities are observed at $\rho\sim1.2\ R_M$.
This region is where the magnetic field is most likely to be connected to the dayside ionosphere, where ionospheric densities are large.
Open and closed magnetic fields anchored to the nightside ionosphere extend tailward within the shadow of the planet, leading to the observed spatial distributions of thermal electrons.
These interpretations are consistent with MHD simulations of the Martian magnetosphere (e.g. \citealt{fang2018}) and its magnetic environment.

The underlying spatial distribution of the magnetic field topology likely plays a significant role in driving the observed trends reported in this paper, in particular with regards to the aforementioned source regions of thermal electrons observed by LPW.
One can imagine an extreme case in which thermal electrons are uniformly distributed throughout the magnetotail region.
Using the definition of ``likelihood'' as defined in our study (occurrence rate of thermal electron detection on a specific magnetic field topology, divided by the total number of thermal electron detections across all topologies), in this extreme case, the likelihood of observing thermal electrons on a specific topology would exactly mirror the likelihood of that topology.
One could redefine the likelihood as the number of thermal electron observations on a specific topology, divided by the total number of observations of that topology (regardless of whether thermal electrons were observed or not).
This analysis is however expected to be subject to additional biases.
For example, upstream solar wind conditions are known to influence magnetotail structure (e.g. \citealt{xu_2020}), and we have subsequently left this analysis for a future comprehensive study of magnetic field topology in the Martian magnetotail.

The locations of the strongest crustal magnetic fields drastically affect the spatial distribution of thermal electrons in the magnetotail region.
Since the strongest crustal fields influence the structure and density of the topside ionosphere \citep{andrews_2015_a}, this result is expected.
Thermal electrons are less likely to be observed in the tail region (within the optical shadow) when the strongest crustal fields are also located on the nightside.
Such conditions suggest that the crustal fields may act to ``trap'' ionospheric thermal electrons at lower altitudes there.
Thermal electrons are also more likely to be observed at $\rho>1\ R_M$ when the strongest crustal fields are on the dayside.
This configuration likely ``puffs up'' the dayside ionosphere (as observed by \citealt{flynn_2017} and \citealt{matta_2015}), resulting in draped and dayside open magnetic fields draping about the terminator region at higher altitudes, as observed in Figure \ref{fig_crustal}~A.


An important caveat to bear in mind when considering the interpretations and implications of this study is that the LPW instrument is only sensitive to electron populations with densities greater than about 15 cm$^{-3}$ and temperatures less than about 1 eV.
As such, the very lowest density populations may not be included here.
Furthermore, the LPW instrument is not capable of determining the bulk flow direction of measured thermal electrons; therefore, we cannot produce an estimate of thermal electron escape rates without assuming a flow direction and speed, which is beyond the scope of this study.


\section{Conclusions} \label{conclude}
This study presents the first detailed analysis of the spatial distribution of thermal (<1 eV) electrons in the Martian magnetotail as observed by the Langmuir Probes and Waves instrument on the MAVEN spacecraft.
The thermal nature of the observed electrons means that they are sourced from the planetary ionosphere.
The presented analysis yields insight into the thermal plasma structure of the Martian magnetotail region and electron source regions.

The spatial distribution of observed thermal electrons varies with both distance down the tail and cylindrical distance from the center of the tail.
We have shown that the local magnetic field flaring direction plays an important role in driving the spatial distribution, with thermal electrons more likely to be observed at greater cylindrical distances for outward flaring field.
Additionally, we have shown that the ionospheric regions to which the local magnetic field connects also drive the spatial distribution of thermal electrons in the magnetotail.
Broadly speaking, thermal electrons observed within the optical shadow behind Mars are typically sourced from the nightside ionosphere, while thermal electrons observed outside of the optical shadow region tend to be sourced from the dayside ionosphere.
When the strongest crustal fields point sunward, thermal electrons are much more likely to be observed in the tail region than when they point tailward.
The observations presented here demonstrate the importance of the magnetic field in structuring the plasma environment of the Martian magnetotail, a characteristic that is likely applicable to other unmagnetized bodies such as Venus and comets.


\clearpage

\acknowledgments
With gratitude, we thank the reviewers who provided valuable feedback on our work.
MN would also like to thank each co-author, Professor David Brain and Dr. R.O. Parke Loyd for their constructive comments and discussions.
Work at LASP and SSL was supported by NASA funding for the MAVEN project through the Mars Exploration Program under grant number NNH10CC04C.
Data used in this study are available on the NASA Planetary Data System, via \href{https://pds.nasa.gov/}{https://pds.nasa.gov/}.

\listofchanges

\end{document}